\documentclass[preprint,eqsecnum,aps,amssymb,tightenlines]{revtex4}

\usepackage{latexsym}

\newcommand{\be}{\begin{equation}}
\newcommand{\ee}{\end{equation}}
\newcommand{\bea}{\begin{eqnarray}}
\newcommand{\eea}{\end{eqnarray}}
\newcommand{\ba}{\begin{array}}
\newcommand{\ea}{\end{array}}
\newcommand{\nn}{\nonumber \\}
\newcommand{\half}{\frac{1}{2}}

\begin{document}

\title{Dispersion Relations in Noncommutative Theories}

\author{Tiago Mariz, J. R. Nascimento}
\affiliation{Departamento de F\'\i sica, Universidade Federal da Para\'\i
  ba \\ 
58051-970, Jo\~ao Pessoa, PB, Brazil \\ 
E-mail: tiago@fisica.ufpb.br, jroberto@fisica.ufpb.br}

\author{Victor O. Rivelles}
\affiliation{Instituto de F\'\i sica, Universidade de S\~ao Paulo\\
 Caixa Postal 66318, 05315-970, S\~ao Paulo, SP, Brazil\\
E-mail: rivelles@fma.if.usp.br} 

\begin{abstract}
We present a detailed study of plane waves in noncommutative abelian
gauge theories. The dispersion relation is deformed from its usual
form whenever a constant background electromagnetic field is
present and is similar to that of an anisotropic medium with no Faraday
rotation nor birefringence.  
When the noncommutativity is induced by the Moyal product we
find that for some values of the background magnetic field no plane
waves are allowed when time is noncommutative. In the Seiberg-Witten
context no restriction is found. We also derive the energy-momentum
tensor in the Seiberg-Witten case. We show that the generalized
Poynting vector obtained from the energy-momentum tensor, the group 
velocity and the wave vector all point in different directions. In the
absence of a constant electromagnetic background we find that the
superposition of plane 
waves is allowed in the Moyal case if the momenta are parallel or
satisfy a sort of quantization condition. We also discuss the relation
between the solutions found in the Seiberg-Witten and Moyal cases
showing that they are not equivalent.

\end{abstract}

\maketitle

\newpage

\section{Introduction}

The fact that coordinates and momenta do not commute in quantum
theory leads naturally to the proposal that coordinates should also be 
noncommuting. This would introduce a new scale in the theory which
could be used to regulate the divergences in quantum field theory
\cite{Snyder:1946qz} but the success of the renormalization program lead
to the dismissal of proposals like this. More recently, however, noncommuting
coordinates were found in several settings involving string theory. In
particular, there is a decoupling limit of open strings in the presence
of D-branes where the effective gauge field theory is defined in a
noncommutative spacetime induced by the Moyal product
\cite{hep-th/9908142} 
\be
\label{Moyal}
A(x) \star B(x) = e^{\frac{i}{2} \theta^{\mu\nu} \partial^x_\mu
  \partial^y_\nu} A(x) B(y)|_{y\rightarrow x},
\ee
where $\theta^{\mu\nu}$ is the noncommutativity parameter \cite{rev}. 
The effect of noncommutativity
in quantum field theory is to add phase factors in the vertices which
produce a mixture of infrared and ultraviolet divergences 
usually breaking down renormalizability \cite{hep-th/9912072}. The only
theories which are known to be free of such a mixing are the
supersymmetric ones \cite{hep-th/0005272}. 

In this context the action for an Abelian gauge field is  
\be
\label{NCaction}
S = -\frac{1}{4} \int d^4x \,\, \hat{F}^{\mu\nu} \star
\hat{F}_{\mu\nu},
\ee
where $\hat{F}_{\mu\nu} = \partial_\mu \hat{A}_\nu - \partial_\nu
\hat{A}_\mu - i [\hat{A}_\mu, \hat{A}_\nu]$ and the brackets denote 
a Moyal commutator. This action is invariant under a non conventional
gauge transformation  
\be 
\delta \hat{A}_\mu = \partial_\mu \hat{\lambda} - i [\hat{A}_\mu,
  \hat{\lambda}]. 
\ee
It is possible to use the Seiberg-Witten map \cite{hep-th/9908142} 
\be
\label{SWmap}
\hat{A}_\mu = A_\mu - \frac{1}{2} \theta^{\alpha\beta}
A_\alpha ( \partial_\beta A_\mu + F_{\beta\mu} ), 
\ee
to get a gauge field $A_\mu$ with the conventional gauge
transformation and an action written in terms of the conventional
field strength. In this picture, the action is expressed as a power
series in the noncommutativity parameter and, to first order in
$\theta$, it is given by 
\be
\label{SW_gauge_action}
S = -\frac{1}{4} \int d^4x \,\,\, \left[ F^{\mu\nu} F_{\mu\nu} + 2
\theta^{\mu\rho} {F_\rho}^\nu \left( {F_\mu}^\sigma F_{\sigma\nu} +
\frac{1}{4} \eta_{\mu\nu} F^{\alpha\beta} F_{\alpha\beta} \right)
\right].
\ee

In the same way that  plane waves can be found in ordinary non-Abelian
gauge theories \cite{Coleman:1977ps} they can also be found in
noncommutative 
theories \cite{hep-th/0006209}. A discussion of waves in more general
noncommutative space-times can be found in
\cite{hep-th/0312328,hep-th/0503087}. Noncommutativity breaks Lorentz
invariance spontaneously due to the existence of a constant matrix
$\theta^{\mu\nu}$ and this means that light waves may no longer travel with
the velocity of light. In the absence of a background electromagnetic field  
the usual dispersion relation is found, whether the
noncommutativity is induced by the Moyal product or by the
Seiberg-Witten map. If a constant electromagnetic background is present the
dispersion relation is changed 
\cite{hep-th/0106044,hep-th/0106047,hep-th/0306272,hep-th/0405253}.  
This clearly opens a new window to detect Lorentz violations effects
due to noncommutativity. 
There are several proposals to find out Lorentz violation and use them
as evidence for quantum gravity effects \cite{gr-qc/0501053}. In
particular, Lorentz violation due to noncommutativity and quantum
gravity effects can be found in the standard model
\cite{Colladay:1998fq}, high energy gamma ray bursts \cite{He:2006yy},
Cerenkov \cite{hep-ph/0411197} and  synchrotron
\cite{Montemayor:2005ka} radiation, among many other examples.  
It is relevant now to study systematically the
modifications induced by noncommutativity in the dispersion
relations and that is the aim of this paper. 

In the next section we will study plane wave solutions and the
corresponding dispersion relations in the Seiberg-Witten map
context. We will obtain the complete dispersion relation when an 
electromagnetic background is present. We find that group and wave
velocities have the same magnitude and that the group velocity is not in the
direction of the wave vector. To find out the direction in which
the energy is being propagated we compute the energy-momentum tensor in Section
III. We choose the energy-momentum tensor which is conserved and gauge
invariant but is neither symmetric nor traceless.  
We then show that the group velocity and the generalized Poynting
vector obtained from the energy-momentum tensor are not in the same
direction. All these effects are similar to those characteristic of an
anisotropic medium. No analogue of Faraday rotation or birefringence
is found since the polarizations travel with the same velocity. 

In Section IV we look for plane wave solutions in the Moyal product
context. We show that there are plane wave solutions
to all orders in $\theta$ and derive the dispersion relation. The
anisotropic effects also show up in this case. Now we
find that the background and the noncommutative are no longer
arbitrary and that there are restriction when the noncommutativity involves
time. In the next section we discuss the equivalence of both 
pictures. We show that plane waves in one picture do not correspond to
plane waves in the other one if the background is the same. We also
show what is the Moyal picture solution corresponding to plane waves
in the Seiberg-Witten context. 

Next we show that two plane waves in the Moyal picture case can obey the
superposition principle if their four-momenta satisfy
$\theta^{\alpha\beta} p_\alpha k_\beta = 2n\pi$, with $n$ an 
integer. They obey the usual dispersion relation. In particular, if the
momenta are in the same direction they can form a wave
packet. Finally, in the last section, we present some conclusions and
further discussions. 

\section{Seiberg-Witten map picture}

In this section we will study some exact solutions to the field
equation coming from the action (\ref{SW_gauge_action}), that is,
\be 
\label{SWfieldeqs}
\partial_\mu F^{\mu\nu} + \theta^{\alpha\beta} {F_\alpha}^\mu (
\partial_\beta {F_\mu}^\nu + \partial_\mu {F_\beta}^\nu ) = 0.
\ee
Clearly, a constant background $F_{\mu\nu}(x) = B_{\mu\nu} =
\mbox{constant}$ is a solution. 
For a plane wave we assume that $F_{\mu\nu}(x) =
\tilde{F}_{\mu\nu}(kx)$. 
Then  the Bianchi identity contracted with $k^\mu$ tell us that 
\be
\label{BI}
k^2 \tilde{F}^\prime_{\mu\nu} + k^\lambda k_\nu
\tilde{F}^\prime_{\lambda\mu} - k^\lambda k_\mu
\tilde{F}^\prime_{\lambda\nu} = 0,
\ee
where $F^\prime$ denotes differentiation with respect to $kx$. Now 
using the field equation (\ref{SWfieldeqs}) we get 
\be
\label{11}
k^2  \tilde{F}^\prime_{\mu\nu} + 2 \theta^{\alpha\beta}
{\tilde{F}_\alpha\,}^\rho k_\beta k_\rho \tilde{F}^\prime_{\mu\nu} = 0.
\ee
The field equation (\ref{SWfieldeqs}) implies that $k^\mu
\tilde{F}^\prime_{\mu\nu}$ is of order $\theta$ and since $F$ and
$F^\prime$ differ by a factor of i, the second term in (\ref{11}) can be
disregarded. Then $k^2 \tilde{F}^\prime_{\mu\nu} = 0$.
We then conclude that for a plane wave the usual dispersion relation
$k^2=0$ holds. After using the Bianchi identity back in the field equation
we get $k^\mu \tilde{F}^\prime_{\mu\nu} = 0$ showing that the plane
wave is transversal like in the commutative case. 

Let us now consider the case of a superposition of a constant
background $B_{\mu\nu}$ and a plane wave
$\tilde{F}_{\mu\nu}(kx)$. Then the field equation (\ref{SWfieldeqs})
becomes 
\be
k_\mu \tilde{F}^{\prime\mu\nu} + \theta^{\alpha\beta} ( {B_\alpha}^\mu
+ {\tilde{F}_\alpha}^{\,\mu} ) ( k_\beta {\tilde{F}_\mu}^{\prime\nu} +
  k_\mu {\tilde{F}_\beta}^{\prime\nu} ) = 0.
\ee
The quadratic terms in $\tilde{F}$ can be disregarded once we use the Bianchi
identities to turn them into the form $k_\mu
{\tilde{F}^{\prime\mu\nu}}$ and then using the fact that it is of order
$\theta$. The equation of motion then reduces to 
\be
\label{fieldeqn}
\tilde{k}_\mu \tilde{F}^{\prime\mu\nu} = 0,
\ee
where
\be 
\label{ktilde}
\tilde{k}_\mu = k_\mu + \theta^{\alpha\beta} B_{\alpha\mu} k_\beta -
      {\theta_\mu}^\alpha{B_\alpha}^\beta k_\beta.
\ee
This is quite interesting since the Bianchi identity is written with
respect to $k_\mu$ as 
\be
\label{Bianchi}
k_\mu \tilde{F}^\prime_{\nu\rho} +k_\nu \tilde{F}^\prime_{\rho\mu} +
k_\rho \tilde{F}^\prime_{\mu\nu}  = 0,
\ee
while the field equation is written with respect to the modified wave
vector $\tilde{k}_\mu$. If we now 
contract the Bianchi identity with $\tilde{k}^\rho$ we get $k_\mu
\tilde{k}^\mu = 0$ or 
\be
\label{DR-SW}
k^2 = - 2 \theta^{\alpha\beta} {B_\alpha}^\rho k_\beta k_\rho.
\ee
Since $k_\mu\tilde{k}^\mu = 0$ we can use (\ref{ktilde}) to get
$\tilde{k}^2 = - k^2$. Notice that (\ref{DR-SW}) is the first sign
that the plane wave velocity may not be equal to the velocity of
light in the presence of a background. 

To solve (\ref{DR-SW})  we take $k^\mu = (\omega, \vec{k})$ so that
$\vec{k}$ can be used to decompose all vectors in components parallel
and perpendicular to it, $\vec{V} = V_L \hat{k} + \vec{V}_T$ 
with $\vec{k} \cdot \vec{V}_T = 0$, and $\hat{k} =
\vec{k}/|\vec{k}|$. We also introduce the vectors 
$\vec{\theta}$ and $\vec{\tilde{\theta}}$ as $\theta^{ij}
= \epsilon^{ijk} \theta^k$ and $\theta^{0i} = \tilde{\theta}^i$,
respectively, and use the vectors $\vec{\cal{E}}$ and $\vec{\cal{B}}$ for the
background $B^{0i} = {\cal{E}}^i$ and $B^{ij} = \epsilon^{ijk} {\cal{B}}^k$,
respectively. An analogous decomposition is used for
$\tilde{F}^{\mu\nu}$. With this notation (\ref{DR-SW}) takes the form 
\begin{equation}
\label{2.9}
\frac{\vec{k}^2}{\omega^2} = 1 - 2 [ \vec{\cal{E}}_T \cdot
  \vec{\tilde{\theta}}_T + \frac{1}{\omega} \vec{k} \cdot
  ( \vec{\cal B}_T \times \vec{\tilde{\theta}}_T) ] - 2 [ \vec{\cal{B}}_T \cdot
  \vec{{\theta}}_T - \frac{1}{\omega} \vec{k} \cdot
  ( \vec{\cal E}_T \times \vec{\theta}_T) ],
\end{equation}
which gives the dispersion relation 
\be
\label{omega}
\omega = | \vec{k} | [ 1 + \vec{\cal{E}}_T \cdot
\vec{\tilde{\theta}}_T + 
\vec{\cal{B}}_T \cdot \vec{{\theta}}_T + \hat{k} \cdot (
\vec{\cal{B}}_T \times \vec{\tilde{\theta}}_T - \vec{\cal{E}}_T \times
\vec{{\theta}}_T ) ].
\ee
This reproduces the results found in
\cite{hep-th/0106044,hep-th/0106047,Chatillon:2006rn,hep-th/0306272}
for several particular cases.

Notice that the frequency is now dependent on the direction of
wave vector, a characteristic of
anisotropic media. We can also compute the
phase and group velocities for each mode. The phase velocity can be
found to be  
\be
\label{vp}
{v}_p = 1 + \vec{\cal E}_T\cdot\vec{\tilde{\theta}}_T + \vec{\cal
  B}_T\cdot\vec{\theta}_T + \hat{k}\cdot( \vec{\cal
  B}_T\times\vec{\tilde{\theta}}_T - \vec{\cal
  E}_T\times\vec{\theta}_T ) = \frac{\omega}{|\vec{k}|},
\ee
and also depends on the wave vector direction. The group velocity is
given by 
\be
\label{vg}
\vec{v}_g = (1 + \vec{\cal E}_T\cdot\vec{\tilde{\theta}}_T + \vec{\cal
  B}_T\cdot\vec{\theta}_T ) \hat{k} - \tilde{\theta}_L \vec{\cal E}_T -
{\cal E}_L \vec{\tilde{\theta}}_T - \theta_L\vec{\cal B}_T - {\cal
  B}_L\vec{\theta}_T + \vec{\cal B}\times\vec{\tilde{\theta}} - \vec{\cal
  E}\times\vec{\theta},
\ee
and it is not in the same direction as the wave vector. It has a component
in the direction of the wave vector which has the same magnitude of
the phase velocity and a transversal component which is first order in
$\theta$. Then, both phase and group velocities have the
same magnitude $v_g = v_p$ to order $\theta$. Notice also that
(\ref{ktilde}) defines a modified wave vector
\be
\label{mwv}
\vec{\tilde{k}} = |\vec{k}| \left[ ( 1 - 2 {\cal
    E}_L\tilde{\theta}_L + 2 \vec{\cal
    B}_T\cdot\vec{\theta}_T ) \hat{k} - \tilde{\theta}_L \vec{\cal 
    E}_T  - {\cal E}_L \vec{\tilde{\theta}}_T  - \theta_L \vec{\cal
    B}_T  - {\cal B}_L \vec{\theta}_T - \vec{\cal E}\times\vec{\theta}
  + \vec{\cal B}\times\vec{\tilde{\theta}} \right].
\ee
We can compute its vector product with the group velocity up to order
$\theta^2$. The result is nonvanishing meaning that the modified wave
vector and the group velocity are not in the same direction. It
remains to be seen whether the plane wave energy is transported along
the direction of $\vec{v}_g$. This will be done in next section.  

Since ${k}_\mu {\tilde{k}}^\mu=0$ the field equation
(\ref{fieldeqn}) reduces to $\tilde{k}_\mu \tilde{A}^\mu = 0$
so that the polarization is orthogonal to $\tilde{k}$. To have
a better understanding of this point let us rewrite the Bianchi
identity (\ref{Bianchi}) in vectorial form as
\bea
\label{1}\vec{k} \cdot \vec{\tilde{B}} = 0, \\
\label{2a} \vec{k} \times \vec{\tilde{E}} - \omega \vec{\tilde{B}} = 0,
\eea
and the field equation (\ref{fieldeqn}) as 
\bea
\label{3} \vec{\tilde{k}} \cdot \vec{\tilde{E}} = 0, \\ 
\label{4} \vec{\tilde{k}} \times \vec{\tilde{B}} + \tilde{\omega}
\vec{\tilde{E}} = 0. 
\eea
From (\ref{1}) we learn that the magnetic field is transversal to
$\vec{k}$ and can be determined by (\ref{2a}) in terms of the
transverse electric field $\vec{\tilde{E}}_T$. Then (\ref{3}) tell us
that the vector field is transverse to $\vec{\tilde{k}}$ so that its
longitudinal component with respect to $\vec{k}$, ${\tilde{E}}_L$,
can be found in terms of $\vec{\tilde{E}}_T$. Finally, (\ref{4}) just
reproduces the dispersion relation $\omega \tilde{\omega} - \vec{k}
\cdot \vec{\tilde{k}} = 0$, so that $\vec{\tilde{E}}_T$ is not
determined. We thus find that the plane wave is transversal and has
two degrees of freedom and both polarizations travel with the same
velocity. 

To untangle the relative directions of the several vectors involved
let us notice that $\vec{\tilde{E}}$ and $\vec{\tilde{B}}$  are
orthogonal to each other and we can use their directions to define two
orthogonal directions. The third orthogonal direction is then defined
by $\vec{\tilde{E}} \times \vec{\tilde{B}}$. Then 
taking the scalar product of $\vec{\tilde{B}}$ with (\ref{2a})
we find that $\vec{k}$ has a component along $\vec{\tilde{E}} \times
\vec{\tilde{B}}$. A similar conclusion holds for
$\vec{\tilde{k}}$. From (\ref{1}) we find that $\vec{k}$ can have a
component along $\vec{\tilde{E}}$ and similarly from (\ref{3})
$\vec{\tilde{k}}$ can have a component along $\vec{\tilde{B}}$.  
Notice that in the pure plane wave case, without any background,
(\ref{1}-\ref{4}) reduce to the same relations found in the absence of
noncommutativity. Then $\vec{\tilde{E}}, \vec{\tilde{B}}$ and
$\vec{k} = \vec{\tilde{k}}$ are mutually orthogonal vectors. 

The next task is to find out the direction in which energy is being
transported. To do so we will compute the energy-momentum tensor. 

\section{The energy-momentum tensor}

The usual properties of the energy-momentum tensor usually do not hold
in noncommutative field theories due to the presence of 
$\theta^{\mu\nu}$. A theory which is invariant under rigid
translations gives rise to a conserved energy-momentum tensor
$T^{\mu\nu}$ which may not be symmetric. However, it can be symmetrized by the
Belinfante procedure.  Lorentz invariance, on the other hand, also
gives rise to a conserved tensor, $M^{\mu\nu\rho}$, such that
$\partial_{\rho} M^{\rho\mu\nu} = T^{\mu\nu} - T^{\nu\mu}$. In a
Lorentz invariant theory $M^{\mu\nu\rho}$ is conserved and
$T^{\mu\nu}$ is symmetric but in noncommutative theories we expected
to find out an antisymmetric part for $T^{\mu\nu}$. Alternatively, we
could enforce a symmetric $T^{\mu\nu}$ in noncommutative theories but
then its conservation is compromised \cite{Das:2002jd}. Also, the
energy-momentum tensor 
obtained before and after the Seiberg-Witten map may not be the same
\cite{Das:2002jd,EMT,Grimstrup:2002xs,Kruglov:2001dm,Banerjee:2003vc}.
Other properties are discussed in \cite{Iorio:2001qy}. 

We are interested in finding the direction where the energy is 
flowing so we need a locally conserved energy-momentum tensor. After the
Seiberg-Witten map, the canonical energy-momentum tensor is 
\cite{Grimstrup:2002xs} 
\be
T_{\mu\nu}^c = 2 {\Pi_\mu}^\alpha \partial_\nu A_\alpha - \eta_{\mu\nu}
{\cal L},
\ee
where $\Pi_{\mu\nu} = \frac{\delta S}{\delta F_{\mu\nu}}$, S is the
action (\ref{SW_gauge_action}) and ${\cal L}$ its Lagrangian. Notice
that $T_{\mu\nu}^c$  is neither symmetric nor traceless. It is
conserved on-shell but it is not gauge invariant. We can apply a sort
of Belinfante procedure \cite{Grimstrup:2002xs} and add a total
derivative to $T_{\mu\nu}^c$ in order to get 
\be
T_{\mu\nu} = 2 {\Pi_\mu}^\alpha F_{\nu\alpha} -  \eta_{\mu\nu}
{\cal L},
\ee
which is also neither symmetric nor traceless, but is gauge
invariant. It is also conserved $\partial_\mu T^{\mu\nu} = 0$ if 
the equations of motion are used. Its explicit form is 
\bea
T_{\mu\nu}& = & \left( 1 - \frac{1}{2} \theta^{\alpha\beta}
F_{\alpha\beta} \right) {F_\mu}^\rho F_{\rho\nu} - {F_\mu}^\alpha
{\theta_\alpha}^\beta {F_\beta}^\gamma F_{\gamma\nu} - {F_\nu}^\alpha
{\theta_\alpha}^\beta {F_\beta}^\gamma F_{\gamma\mu} - \nn
& & - {\theta_\mu}^\alpha {F_\alpha}^\beta {F_\beta}^\gamma F_{\gamma\nu} 
- \frac{1}{4} {\theta_\mu}^\alpha F_{\alpha\nu} F^2 - \eta_{\mu\nu}
{\cal L},
\eea
and agrees with the results of
\cite{Grimstrup:2002xs,Kruglov:2001dm}. After a lengthy calculation we
can find its components 
\bea
\label{00}
T^{00} &=& \frac{1}{2} ( 1 + \vec{\theta} \cdot \vec{B} )
( \vec{E}^2 + \vec{B}^2 ) -
( \vec{\tilde{\theta}}\cdot\vec{E} )  \vec{E}^2 -
( \vec{\theta}\cdot\vec{E} ) (\vec{E}\cdot\vec{B}) , \\
\label{i0}
T^{0i} &=& ( 1 - \vec{\tilde{\theta}}\cdot\vec{E} +
  \vec{\theta}\cdot\vec{B} ) (\vec{E}\times\vec{B})^i -
\frac{1}{2} (\vec{E}^2 - \vec{B}^2)
(\vec{\tilde{\theta}}\times\vec{B})^i - (\vec{E}\cdot\vec{B})
(\vec{\theta}\times\vec{B})^i, \\
T^{i0} &=& ( 1 - \vec{\tilde{\theta}}\cdot\vec{E} +
  \vec{\theta}\cdot\vec{B} ) (\vec{E}\times\vec{B})^i +
\frac{1}{2} (\vec{E}^2 - \vec{B}^2)
(\vec{\theta}\times\vec{E})^i - ( \vec{E}\cdot\vec{B} )
(\vec{\tilde{\theta}}\times\vec{E})^i, \\
T^{ij} &=& - ( 1 + \vec{\theta}\cdot\vec{B} -
\vec{\tilde{\theta}}\cdot\vec{E} ) ( E^i E^j + B^i B^j) + \nn
& & 
+ \frac{1}{2} (\vec{E}^2 - \vec{B}^2 ) ( 
\tilde{\theta}^i E^j + B^i \theta^j ) - 
 ( \vec{E}\cdot\vec{B}) (B^i
\tilde{\theta}^j - \theta^i E^j ) +
\nn
& & +  \delta^{ij} \left[ \frac{1}{2} (1
  - \vec{\tilde{\theta}}\cdot\vec{E}) (\vec{E}^2 + \vec{B}^2) +
  ( \vec{\theta}\cdot\vec{B} ) \vec{B}^2 +
  ( \vec{\tilde{\theta}}\cdot\vec{B} ) (\vec{E} \cdot\vec{B}) \right].
\eea
Some pieces were already known in particular cases. For instance, when
$\vec{\tilde{\theta}}=0$, (\ref{00}) agrees with the result in
\cite{Kruglov:2001dm}.  

The presence of an antisymmetric part in the energy-momentum tensor
requires some care. We can still interpret $T^{i0}$ as a sort
of generalized Poynting vector and $T^{00}$ as an energy density
because $T^{\mu\nu}$ is locally conserved. 
Notice that $T^{00}$ does not seem to be positive definite. The
noncommutative contributions  
proportional to $\vec{E}^2$ and $\vec{B}^2$ are harmless because
$\theta^{\mu\nu} F_{\mu\nu} << 1$ and $\frac{1}{2}(\vec{E}^2 +
\vec{B}^2)$ is larger than them. The noncommutative term with
$\vec{E}\cdot\vec{B}$ is also small than the commutative term
$\frac{1}{2}(\vec{E}^2 + \vec{B}^2)$ because $\vec{E}\cdot\vec{B} \le
\frac{1}{2}(\vec{E}^2 + \vec{B}^2)$. 
So, somewhat surprisingly, the energy density (\ref{00}) is positive
definite for small noncommutativity. 

To find out the direction of the energy flux let us manipulate 
(\ref{1}-\ref{4}). We can use (\ref{2a}) to find that
\be
\label{order}
\vec{\tilde{B}}^2 = \frac{\vec{k}^2}{\omega^2} \vec{\tilde{E}}^2 -
\frac{1}{\omega^2} (\vec{k}\cdot\vec{\tilde{E}})^2.
\ee
By (\ref{2.9}) we know that
$\vec{k}^2/\omega^2 = 1 + {\cal O}(\theta)$. From (\ref{4}) we
find that $\vec{k}\cdot\vec{\tilde{E}} = -\frac{1}{\tilde{\omega}}
\vec{\tilde{B}}\cdot(\vec{k}\times\vec{\tilde{k}})$, but from (\ref{mwv}) we
get that $\vec{k}\times\vec{\tilde{k}}$ is of order $\theta$ and so is
$\vec{k}\cdot\vec{\tilde{E}}$. Then, from (\ref{order}) we find that
also $\vec{\tilde{E}}^2 - \vec{\tilde{B}}^2$ is of order $\theta$. 

Consider first the case of vanishing background. Since 
$\vec{\tilde{E}}$ and $\vec{\tilde{B}}$ are orthogonal to each other
and $\vec{\tilde{E}}^2 - \vec{\tilde{B}}^2 = 0$ only the first term of
$T^{i0}$ contributes and the energy flux is in the direction of
$\vec{\tilde{E}}\times\vec{\tilde{B}}$. We can now take a time
average and only the quadratic terms will survive. This means that all
noncommutative contributions vanish and we get the commutative
Poynting vector as a result. We can also take the time average of
$T^{00}$. All noncommutative contributions are cubic in the fields and
vanish. We get the same energy density as in the commutative case. It
is quite interesting that in the absence of a background the 
noncommutative plane wave behaves like in the commutative case. 

Let us return to the case where the background is present. Now
$\vec{E}^2 - \vec{B}^2$ is no longer of order $\theta$ but
proportional to the background fields and the plane wave. Also,
$\vec{E}\cdot\vec{B}$ no longer vanishes because of the background
contribution. So $T^{i0}$ will in general have all terms present. Even
if we take a time average many terms will survive. This means that the
direction of the energy flux will be the direction
$\vec{E}\times\vec{B}$ plus small noncommutative corrections. Notice
also that both $\vec{E}$ and $\vec{B}$ depend on the background so the
direction of the energy flux will be background dependent. 

We can now check whether the energy flux is in the direction of the group
velocity. We can take the vector product of the time averaged $T^{i0}$
with the group velocity to order $\theta^2$ and verify that it does not 
vanish. Therefore, the direction of the Poynting vector and the group
velocity do not coincide. Also, the vector product with either the wave
vector $\vec{k}$ or the modified wave vector $\vec{\tilde{k}}$ does not
vanish confirming the anisotropic properties produced by the
background. Since the polarizations travel with the same 
velocity neither Faraday rotation nor birefringence is present. 

\section{Moyal product picture}

In the Moyal product picture the field equation derived from
(\ref{NCaction}) is
\be
\label{NCfieldeqs}
\hat{D}_\mu \hat{F}^{\mu\nu} = \partial_\mu \hat{F}^{\mu\nu} - i
    [\hat{A}_\mu, \hat{F}^{\mu\nu}] =  0.
\ee
The solution for a constant background is \cite{hep-th/0006209} 
\be
\label{constantsolution}
\hat{A}_\mu = -\half B_{\mu\nu} x^\nu,
\ee
with $B_{\mu\nu}$ again constant. Notice that $B_{\mu\nu}$ does not
need to be antisymmetric. The field strength is given by
\be
\hat{F}_{\mu\nu} = B_{\mu\nu} + \frac{1}{4} \theta^{\alpha\beta}
B_{\mu\alpha} B_{\nu\beta},
\ee
and it satisfies the field equation (\ref{NCfieldeqs}) to all
orders in $\theta$. Notice also that the field strength can vanish by an
appropriate choice of the background. 

For a plane wave we choose \cite{hep-th/0006209}
\be
\label{planewave}
\hat{A}_\mu(x) = \tilde{A}_\mu (kx),
\ee
and we find that $\hat{F}_{\mu\nu} = k_\mu \tilde{A}^\prime_\nu -
k_\nu \tilde{A}^\prime_\mu$ to all orders in $\theta$ since the
commutator term in (\ref{NCfieldeqs}) does not give any 
contribution. The field equation then reads
\be
k^2 \tilde{A}^{\prime\mu} - k^\mu k_\nu \tilde{A}^{\prime\nu} = 0,
\ee
and we find a solution if $ k^2 = 0$ and $k^\mu
\hat{A}_\mu = 0$. Then a transversal plane wave is also a solution to
all orders in $\theta$. 

Remarkably, the superposition of a constant background
(\ref{constantsolution}) and a plane wave (\ref{planewave}) also 
constitutes a solution to all orders. To show this we first notice 
that the field strength is given to all orders in $\theta$ by 
\be
\label{FE}
\hat{F}_{\mu\nu} = B_{\mu\nu} + \frac{1}{4} \theta^{\alpha\beta}
B_{\mu\alpha} B_{\nu\beta} + \overline{k}_\mu \tilde{A}^\prime_\nu - 
\overline{k}_\nu\tilde{A}^\prime_\mu,
\ee
where 
\be
\label{kbar}
\overline{k}_\mu = k_\mu - \half \theta^{\alpha\beta} B_{\mu\alpha} 
k_\beta.
\ee
This means that the effect of the background on the wave vector is to
replace it by $\overline{k}^\mu$. 
Now, by applying the covariant derivative $\hat{D}^\rho$ to the Bianchi
identity 
\be
\label{BianchiNC}
\hat{D}_\rho \hat{F}_{\mu\nu} + \hat{D}_\nu \hat{F}_{\rho\mu} +
\hat{D}_\mu \hat{F}_{\nu\rho} = 0,
\ee
and using the equation of motion (\ref{NCfieldeqs}) we find 
\be
\hat{D}^2 \hat{F}_{\mu\nu} - i [ {\hat{F}_\mu\,}^\rho, \hat{F}_{\rho\nu}
] +  i [ {\hat{F}_\nu\,}^\rho, \hat{F}_{\rho\mu} ] = 0.
\ee
For our solution we find, using (\ref{FE}), that the commutator terms
vanish so that $\hat{D}^2 \hat{F}_{\mu\nu} = 0$.

On the other side, taking the covariant derivative of (\ref{FE}) we
find to all orders in $\theta$ that 
\be
\label{fieldeqn1}
\hat{D}_\rho \hat{F}_{\mu\nu} = \overline{k}_\rho \tilde{F}_{\mu\nu},
\ee
where $\tilde{F}_{\mu\nu} = \overline{k}_\mu \tilde{A}^\prime_\nu - 
\overline{k}_\nu \tilde{A}^\prime_\mu$, so that $\hat{D}^2
\hat{F}_{\mu\nu} = \overline{k}^2  \tilde{F}_{\mu\nu}$. Taking into
account that $\hat{D}^2 \hat{F}_{\mu\nu} = 0$ we find that  
$\overline{k}^2 = 0$ or
\be
\label{DR-MP}
k^2 = 2k_\mu V^\mu - V_\mu V^\mu,
\ee
where
\be
\label{V}
V_\mu = \half \theta^{\alpha\beta} B_{\mu\alpha} k_\beta.
\ee

Going back to the field equation we find, using (\ref{fieldeqn1}), that 
\be
\hat{D}_\mu \hat{F}^{\mu\nu} = - \overline{k}^\nu \overline{k}_\mu
\tilde{A}^{\prime\mu},  
\ee
so that $\overline{k}_\mu \tilde{A}^\mu = 0$. 

Let us now focus on the plane wave contribution. Its field strength is
given by 
$\tilde{F}_{\mu\nu} = \overline{k}_\mu \tilde{A}^\prime_\nu -  
\overline{k}_\nu \tilde{A}^\prime_\mu$ and taking into account
(\ref{fieldeqn1}) it satisfies
\be
\hat{D}_\rho \tilde{F}_{\mu\nu} = \overline{k}_\rho \tilde{F}_{\mu\nu}.
\ee
This means that the Bianchi identity (\ref{BianchiNC}) for
$\tilde{F}$ now reduces to 
\be
\overline{k}_\mu \tilde{F}^\prime_{\nu\rho} + \overline{k}_\nu
\tilde{F}^\prime_{\rho\mu} + \overline{k}_\rho
\tilde{F}^\prime_{\mu\nu}  = 0,
\ee
while the equation of motion becomes $\overline{k}_\mu
\tilde{F}^{\mu\nu} = 0$. This means that the  electric and magnetic
components are orthogonal to $\vec{\overline{k}}$ and not to
$\vec{k}$. 

To find the dispersion relation from (\ref{DR-MP}) we must first notice
that it gives a second degree equation for $\omega$. This
means that the solution will depend on the value of the discriminant
\be
\label{disc}
\Delta = (2 \alpha \vec{\tilde{k}} - V^0 \vec{b})^2 - |
\vec{\overline{k}} \times \vec{b}|^2,
\ee
where 
\be 
\vec{b} = \vec{\tilde{\theta}} \times \vec{\cal B}, \quad 
\alpha = 1 - \half \vec{\cal E} \cdot \vec{\tilde{\theta}}.
\ee
We could not find a closed form for a generic value of $\Delta$ so 
we will analyze the possible solutions according to the
noncommutativity which is present. Since the second degree equation
has in general two solutions we choose the one which reproduces the
usual dispersion relation in the commutative limit. 

In the noncommutative magnetic case, that is when 
$\vec{\tilde{\theta}}=0$, the discriminant (\ref{disc}) is always
positive and it is easy to find the solution
\be
\omega = |\vec{k}| \left( | (1 + \frac{1}{2} \vec{\cal
    B}_\cdot\vec{\theta} ) \hat{k} - \frac{1}{2}
  (\hat{k}\cdot\vec{\cal B}) \vec{\theta} | - \frac{1}{2}
  \hat{k}\cdot(\vec{\cal E}\times\vec{\theta}) \right).
\ee
This result agrees with \cite{Chaichian:2005gh}. 

In the electric case, $\vec{\theta}=0$, the
discriminant is not positive definite so we have to consider the
effect of the background. If the background is purely electric,
$\vec{\cal B}=0$, then the discriminant is always positive and we have
\be
\label{electriccase1}
\omega = |\vec{k}| \frac{ | \hat{k} - \half (\vec{k} \cdot
  \vec{\tilde{\theta}} ) \vec{\cal E} | }{1 - \half
  \vec{\tilde{\theta}} \cdot \vec{\cal E} }.
\ee
In the case where the background is purely magnetic, $\vec{\cal E}=0$, we
  find
\be
\omega = \frac{|\vec{k}|}{\left( 1 - \frac{1}{4} ( \vec{\tilde{\theta}}
  \times \vec{\cal B} )^2 \right) } \left[  \sqrt{ 1 - \frac{1}{4} 
|\hat{k} \times ( \vec{\tilde{\theta}} \times \vec{\cal B} ) |^2 } 
- \hat{k} \cdot ( \vec{\tilde{\theta}} \times \vec{\cal B} ) \right], 
\ee
and the plane wave solution exists only when $1 - \frac{1}{4} |
\hat{k} \times ( \vec{\tilde{\theta}} \times \vec{\cal B} ) |^2 \ge
0$. Then, in the case of space-time noncommutativity we find that
there is a restriction for the existence of plane waves. It is curious
that precisely for this case the quantum theory is problematic since unitarity
is lost \cite{Gomis:2000zz}. In the other cases there are no
restriction for the existence of plane waves. Notice also that all
dispersion relations depend on the wave vector direction so in all
cases the presence of an electromagnetic background simulates an
anisotropic medium.  

\section{Equivalence of both pictures}

The Seiberg-Witten map (\ref{SWmap}) is a change of variables which
preserves the gauge orbits. Since the physics can not depend on the
choice of variables we expect that the results obtained in the two
pictures should be equivalent. However, taking the noncommutative
dispersion relation (\ref{DR-MP}) to order $\theta$, and assuming that
the background is the same in both cases, we get $ k^2 = 2
k_\mu V^\mu$, and not $k^2 = 4 k_\mu V^\mu$ as required by
(\ref{DR-SW}). We can also take  
the Moyal picture solution of a constant background
and a plane wave (\ref{FE}) and apply the Seiberg-Witten map to
it. The resulting field strength is not of the form $B_{\mu\nu} +
\tilde{F}_{\mu\nu} (kx)$ but has extra pieces linear in
$x^\mu$ so it does not correspond to a superposition of a background
with a plane wave in the Seiberg-Witten picture. Hence, the field
configurations are not equivalent. Since 
$F_{\mu\nu}$ is gauge invariant the terms linear in $x^\mu$ can not be
removed by a gauge transformation. This shows that the solutions are
inequivalent and it does not make any sense to try to compare the
results in different pictures. 

However we can find the solution in the Moyal picture which
is equivalent to the superpositions of a background plus a plane
wave in the Seiberg-Witten picture. It is given by  
\be
\label{NC1}
\hat{A}_\mu(x) = \tilde{A}_\mu(kx) - V_\nu x^\nu \tilde{A}_\mu(kx) -
\frac{1}{2} \hat{B}_{\mu\nu} x^\nu, 
\ee
with $V_\mu$ given by (\ref{V}). Using the Seiberg-Witten map we get 
\be
\label{C1}
A_\mu = \tilde{A}^{(c)}_\mu (kx) - \frac{1}{2} B_{\mu\nu} x^\nu,
\ee
where
\bea
\tilde{A}^{(c)}_\mu (kx) &=& [ 1 + \theta^{\alpha\beta} \tilde{A}_\alpha
(kx) k_\beta ] \tilde{A}_\mu (kx) + \theta^{\alpha\beta}
\tilde{A}_\alpha(kx) \hat{B}_{\beta\mu}, \nn
B_{\mu\nu} &=& \hat{B}_{\mu\nu} - \frac{3}{4} \hat{B}_{\mu\alpha}
\hat{B}_{\nu\beta}.
\eea
This means that $F_{\mu\nu}$ has the form  $B_{\mu\nu} +
\tilde{F}_{\mu\nu} (kx)$ and describes a plane wave. Notice that
$\hat{A}_\mu$ and $A_\mu$ in 
(\ref{NC1}) and (\ref{C1}), respectively, are related by the
Seiberg-Witten map if the momentum is also transformed as
$\hat{k}^\mu = k^\mu - V^\mu$. Now we get the correct dispersion
relation and polarization condition for $\tilde{A}_\mu$ in the Seiberg-Witten
picture. 

\section{Superposition of plane waves}

We now consider the superposition of two plane waves with different momenta
\be
\label{2planewaves}
\hat{A}_\mu (x) = \hat{A}_{1\mu} (kx) + \hat{A}_{2\mu} (px), \quad
p^\mu \not= k^\mu.
\ee
The field strength is easily found to be
\be
\hat{F}_{\mu\nu} = k_\mu \hat{A}_{1\nu} - k_\nu \hat{A}_{1\mu} + p_\mu
\hat{A}_{2\nu} - p_\nu \hat{A}_{2\mu} + 2 \sin(\frac{k\theta p}{2} ) (
\hat{A}_{1\mu} \hat{A}_{2\nu} - \hat{A}_{1\nu} \hat{A}_{2\mu} ),
\ee
where $k\theta p = k_\alpha \theta^{\alpha\beta} p_\beta$. The
equation of motion takes the form 
\bea
&& k_\mu k^{[\mu} \hat{A}_1^{\nu]} + p_\mu p^{[\mu} \hat{A}_2^{\nu]} + 2
\sin(\frac{k\theta p}{2} ) ( k_\mu + p_\mu) \hat{A}_1^{[\mu}
\hat{A}_2^{\nu]} + 2 \sin(\frac{k\theta p}{2} ) \left( \hat{A}_{1\mu}
  p^{[\mu} \hat{A}_2^{\nu]} - \hat{A}_{2\mu} k^{[\mu} \hat{A}_1^{\nu]}
  + \right.  \nn 
&& \left. 2 \sin(\frac{k\theta p}{2} ) ( \hat{A}_{1\mu}
- \hat{A}_{2\mu} ) \hat{A}_1^{[\mu} \hat{A}_2^{\nu]} \right) = 0, 
\eea
and it is easily seen that there is a non-trivial solution if 
\be
\label{sup}
k\theta p = 2n\pi, \quad k^2 = p^2 = 0, \quad k_\mu \hat{A}_1^\mu =
p_\mu \hat{A}_2^\mu  = 0,
\ee
where $n$ is an integer. This shows that it is possible to have a
superposition of two transversal plane waves in a noncommutative
theory if the wave vectors $k^\mu$ and $p^\mu$ are parallel or satisfy
$k\theta p = 2n\pi$. The dispersion relation is the same as in
the commutative case. In fact, this can be easily generalized to a
finite number of plane waves. 

In the Seiberg-Witten picture there is no solution corresponding to a
superposition of plane waves. This is due to the nonlinear terms
present in the field equation (\ref{SWfieldeqs}).

\section{Conclusions}

We have shown the existence of plane wave solutions in noncommutative
abelian gauge theories. In both pictures they present a
deformed dispersion relation in the presence of a electromagnetic
background.  The dispersion relation depends on the wave vector
direction and presents similar properties to those found when we
consider the propagation of light in an anisotropic medium. It is worth
noticing that if $\tilde{\theta}$ is not vanishing, that is, when the
noncommutativity involves time, there are restrictions on the
background for the existence of plane wave solutions in the Moyal
picture but not in the Seiberg-Witten one. Remarkably, the
Moyal picture allows solutions involving a superposition of 
plane waves. In this case the momenta are either parallel or satisfy
(\ref{sup}). 

In the Seiberg-Witten picture we also discussed the energy-momentum
tensor. It can be used to define a generalization of the Poynting
vector and energy density to the noncommutative case. The Poynting
vector, the group velocity, the wave vector and the modified wave
vector all point in different directions. Even so, the generalized
Poynting vector represents the transport of energy since it obeys a
continuity equation. This means that the effect of the background
electromagnetic field in the presence on noncommutativity can be
interpreted as an anisotropic medium which presents neither Faraday
rotation nor birefringence effects. 

We also showed that plane waves in one picture does not correspond to
plane waves in the other picture. This means that extreme care must be
taken when comparing results in different pictures. 
Since there are many proposed tests for Lorentz violation in several
settings it is very important to understand the noncommutative
contribution to them. The results presented here are just the first
steps in this direction.

\section{Acknowledgements}

The work of TM and JRN was supported by CAPES, CNPq, PADCT/CNPq, and
PRONEX/CNPq/FAPESQ. The work of VOR was supported by CNPq, FAPESP and
PRONEX under contract CNPq 66.2002/1998-99.

\end{document}